\documentstyle[aps]{revtex}
\begin{document}

\draft

\title{Entanglement measure and distance}
\author{Fumiaki Morikoshi
\thanks{{\ttfamily morikosh@particle.sci.hokudai.ac.jp}}}
\address{Department of Physics\\ Hokkaido University\\
         Sapporo, 060-0810, Japan}

\maketitle

\begin{abstract}

Recently a new entanglement dilution scheme
has been constructed by Lo and Popescu.
This paper points out that this result has a deep implication that
the entanglement measure for bipartite pure states is independent of
the distance between entangled two systems.

\end{abstract}


Recent developments in quantum information theory show that
quantum entanglement realizes novel communication,
like quantum teleportation \cite{Bennett1}
and superdense coding \cite{Bennett2},
that is impossible in a classical manner.
A key of quantum information processing is
how we exploit quantum entanglement.
Though entanglement is a valuable resource,
we have not yet uncovered its properties completely.
Thus one of the central tasks of quantum information theory
is to characterize and quantify entangled states.

In the case of bipartite pure entangled states,
a unique entanglement measure, entropy of entanglement,
has already been established \cite{Bennett3,Lo1}.
We can quantify various bipartite pure entangled states
by this entanglement measure.
An important feature of quantum entanglement is that
its physical properties do not depend on the distance between
entangled two systems.
How far the two systems are apart does not concern
to what extent they are entangled.
Therefore the entanglement measure for bipartite pure states
should not depend on the distance between them.
In the following, we will discuss this problem in detail.

In this paper we consider only bipartite pure entangled states.
Although we deal with only entangled states of two qubits
to make the following argument clear,
the result discussed below is easily generalized to bipartite pure-state
entanglement of higher dimensional states.

An arbitrary bipartite pure state is described by Schmidt decomposition as
\begin{equation}
 |\psi \rangle _{AB} = a |00 \rangle _{AB} + b |11 \rangle _{AB},
\label{eq:Schmidt}
\end{equation}
where $a$ and $b$ are real non-negative numbers and
satisfy a normalization condition $a^{2} + b^{2} = 1$.
One of the most useful entangled state is a Bell pair
\begin{equation}
 |\phi^{+} \rangle _{AB} = \frac{1}{\sqrt{2}} (|00 \rangle _{AB}
                          + |11 \rangle _{AB}).
\label{eq:Bell}
\end{equation}
Suppose two qubits A and B are possessed by Alice and Bob, respectively.
By using a Bell pair, they can perform novel communication.
For example, in quantum teleportation
Alice can send an unknown qubit state to Bob only by local operations
and classical communication with the assistance of a Bell pair.
In superdense coding,
Alice can send two bits to Bob by performing local operations and
sending only one qubit if they share a Bell pair in advance.

If the entangled state shared between Alice and Bob is not a Bell pair,
but a partially entangled pair represented by eq.\ (\ref{eq:Schmidt}),
these quantum communication can not be performed with perfect fidelity.
Thus Alice and Bob must convert partially entangled states
[eq.\ (\ref{eq:Schmidt})] into Bell pairs [eq.\ (\ref{eq:Bell})]
by local operations and classical communication.

The most effective conversion procedure is a collective manipulation of
many copies of the state $|\psi \rangle_{AB}$,
not an individual manipulation of each pair.
We can quantify the number of Bell pairs obtained from $|\psi \rangle_{AB}$
by the entropy of entanglement $E(|\psi \rangle_{AB})$ defined as
\begin{equation}
E(|\psi \rangle_{AB}) = -a^{2} \log_{2} a^{2} -b^{2} \log_{2} b^{2},
\label{eq:entropy}
\end{equation}
which is an intrinsic quantity to $|\psi \rangle_{AB}$
[eq.\ (\ref{eq:Schmidt})].
It is proved that $n$ copies of the state $|\psi \rangle_{AB}$
can be converted into $nE(|\psi \rangle_{AB})$ Bell pairs
in the asymptotic limit $n \to \infty$
and no more Bell pairs can be obtained \cite{Bennett3,Lo1}.
In other words, $(|\psi \rangle_{AB})^{n}$ can be converted into
$(|\phi^{+} \rangle_{AB})^{nE}$ asymptotically.
This conversion procedure is called entanglement concentration.
Though the concrete procedure of entanglement concentration is not
illustrated here,
it is important that the concentration process needs no classical
communication
between Alice and Bob.

On the other hand, if Alice and Bob share $nE(|\psi \rangle_{AB})$ Bell
pairs,
they can convert $(|\phi^{+} \rangle_{AB})^{nE}$ into
$(|\psi \rangle_{AB})^{n}$ in the $n \to \infty$ limit.
They can not obtain $(|\psi \rangle_{AB})^{n}$ from less Bell pairs.
This procedure is called entanglement dilution.
The concrete dilution procedure is described as follows \cite{Bennett3}.
First Alice prepares $n$ copies of the state
$|\psi \rangle _{AC}$ locally.
Then she Schumacher compresses the system $C$ \cite{Schumacher},
which reduces the number of qubits
containing meaningful quantum information in the system $C$
from $n$ to $nE(|\psi \rangle _{AC})$.
Next she teleports these $nE(|\psi \rangle _{AC})$ qubits to Bob
using previously shared $nE(|\psi \rangle _{AC})$ Bell pairs
i.e., $(|\phi^{+} \rangle_{AB})^{nE}$.
Finally Bob Schumacher decompresses the system $B$
that is now in the same state as the system $C$ was before teleportation.
In the asymptotic limit,
this process successfully produces the state
$(|\psi \rangle _{AB})^{n}$
between Alice and Bob with arbitrarily good fidelity.
However, this dilution process contains classical communication
because of quantum teleportation.
In this dilution scheme,
Alice and Bob need 2 bits of classical communication per Bell pair.

Alice and Bob can concentrate $n$ pairs of $|\psi \rangle _{AB}$
into $nE(|\psi \rangle _{AB})$ Bell pairs and
dilute $nE(|\psi \rangle _{AB})$ Bell pairs into
$n$ pairs of $|\psi \rangle _{AB}$ conversely.
This means that these two operations are reversible
in the sense that these can be performed without any loss of entanglement.
Thus it is concluded that the state $|\psi \rangle _{AB}$ is
equivalent to $E(|\psi \rangle _{AB})$ Bell pairs in the asymptotic limit.
This fact enables us to quantify arbitrary entangled states
by the entropy of entanglement $E(|\psi \rangle _{AB})$
[eq.\ (\ref{eq:entropy})].
The unit of this measure is a Bell pair i.e.,
$E(|\phi^{+} \rangle_{AB}) = 1$.
Eq.\ (\ref{eq:entropy}) is a special case for entangled states
consisting of two qubits.
In general, the entanglement measure for bipartite pure states
is the von Neumann entropy of Alice's (or Bob's) density matrix i.e.,
\begin{equation}
E(|\psi \rangle_{AB}) = -tr(\rho_{A} \log_{2} \rho_{A}),
\end{equation}
where $\rho_{A} = tr_{B} (|\psi \rangle_{AB AB} \langle \psi|)$.
The reversibility between concentration and dilution is crucial to
the foundation of this entanglement measure.
It has been proved that this is a unique measure for bipartite
pure-state entanglement \cite{Popescu}.

It is believed that the above argument establishes the entanglement measure
for bipartite pure states.
Indeed, any entangled states can be quantified by this measure.
However, the above argument is insufficient to define the entanglement
measure
independent of the distance between Alice and Bob.
The entanglement measure should have the same feature
as quantum entanglement that physical properties are independent of
the distance between Alice and Bob.
In the following, we will see that the classical communication
in dilution processes places a certain restriction on the reversibility
between concentration and dilution and thus causes the dependence on
the distance.

Though the entanglement measure represented by eq.\ (\ref{eq:entropy})
does not contain the distance between Alice and Bob explicitly,
the reasoning that defines the measure involves the distance
through classical communication.
When we concentrate partially entangled states,
we need no classical communication in the asymptotic limit,
but in diluting Bell pairs, we need 2 bits of classical communication
per Bell pair.
This means that the reversible cycle of concentration and dilution
requires 2 bits of classical communication per Bell pair
even in the asymptotic limit.
The entanglement measure $E(|\psi \rangle_{AB})$
has its foundation on the fact that $|\psi \rangle_{AB}$ and
$E(|\psi \rangle_{AB})$ Bell pairs are reversibly convertible each other.
Therefore the definition of the entanglement measure must be
associated with classical communication,
which raises a subtle problem.
When Alice and Bob convert the state $|\psi \rangle_{AB}$
through the reversible cycle,
classical communication cost they need is dependent on the distance
between them.
For example, let us compare two cases where the distance between
them are $d$ and $2d$.
In the latter case, the dilution part in the reversible conversion cycle
takes twice as long time as in the former case
because of classical communication.
As far as the classical communication cost
dependent on the distance is concerned,
more cost is needed in the latter than in the former.
In order to define the entanglement measure in the distance independent
manner,
the property of the reversible cycle must be independent of the distance.
Thus the entanglement measure is not independent of the distance
between Alice and Bob for the reason stated above.

Recently Lo and Popescu have presented a new entanglement
dilution procedure that can be performed with vanishing classical
communication cost per Bell pair in the asymptotic limit \cite{Lo2}.
By the new dilution scheme, they concluded that
entanglement is truly interconvertible resource and that
``ebit'' (the unit of entanglement) is established independent of ``bit''.

However, from the above argument we can point out that
their new dilution scheme has a deeper implication.
Replacing the ordinary dilution scheme with the new one,
Alice and Bob can convert the entangled states
with vanishing classical communication cost per Bell pair
in the asymptotic limit.
Since the reversibility between concentration and dilution
is not associated with classical communication cost anymore,
a definite amount of entanglement can be reversibly converted
in the distance independent manner.
Thus we can conclude that the entanglement measure for bipartite pure states
is truly independent of the distance between Alice and Bob.

In the case of multipartite pure-state entanglement,
it is easily understood that the Schmidt decomposable states such as
\begin{equation}
|\Psi \rangle_{AB \cdots N} = a|00 \cdots 0 \rangle_{AB \cdots N}
                             +b|11 \cdots 1 \rangle_{AB \cdots N},
\end{equation}
can be treated in the same way as in the bipartite case.
The n-qubit cat state
\begin{equation}
|cat \rangle_{AB \cdots N} = \frac{1}{\sqrt{2}}
                             (|00 \cdots 0 \rangle_{AB \cdots N}
                              +|11 \cdots 1 \rangle_{AB \cdots N}),
\end{equation}
plays a similar role to a Bell pair.
However, the entanglement measure for general multipartite pure states
has not been established.
Thus, whether the entanglement measure for multipartite pure states
is also independent of the distances between all entangled systems or not
remains as an open question.

\acknowledgments

I am grateful to K.Suehiro
for careful reading of the manuscript and helpful comments.
I am also indebted to N.Imoto and H.-K. Lo for pointing out
the fault in the previous version.

\end{document}